\documentclass[aps,showpacs]{revtex4}
\newcommand{\nonum}{\nonumber\\}
\newcommand{\ba}{\begin{array}}
\newcommand{\ea}{\end{array}}
\newcommand{\beq}{\begin{equation}}
\newcommand{\eeq}{\end{equation}}
\newcommand{\bea}{\begin{eqnarray}}
\newcommand{\eea}{\end{eqnarray}}
\newcommand{\beal}{\setcounter{letter}{1} \begin{eqnarray}}
\newcommand{\eeal}{\addtocounter{equation}{1} \end{eqnarray}}

\newcommand{\larrow}{\,\,\,\,\hbox to 30pt{\rightarrowfill}
\,\,\,\,}
\newcommand{\slarrow}{\,\,\,\hbox to 20pt{\rightarrowfill}
\,\,\,}

\newcommand{\IR}{{\rm I\kern-.22em R}}
\begin{document}
\title{\bf  $d$-Dimensional Black Hole Entropy Spectrum from Quasi-normal Modes}

\author{G. Kunstatter}
\affiliation{Winnipeg Institute for Theoretical Physics}
\affiliation{Physics Dept. University of Winnipeg, Winnipeg, Manitoba CANADA R3B 2E9 }

\date{\today}

\begin{abstract}
Starting from recent observations\cite{hod,dreyer1} about quasi-normal modes,
we use semi-classical arguments to derive the Bekenstein-Hawking entropy spectrum for $d$-dimensional spherically symmetric black holes .
We find that, as first suggested by Bekenstein, the entropy spectrum is equally spaced:
$S_{BH}=k \ln(m_0) n$, where $m_0$ is a fixed
integer that must be derived from the microscopic theory. As shown in \cite{dreyer1},4-$d$ loop quantum gravity  yields precisely such a spectrum with $m_0=3$ providing the Immirzi parameter is chosen appropriately. For $d$-dimensional black holes of radius $R_H(M)$, our analysis 
predicts 
the existence of a unique quasinormal mode frequency in the large damping limit
$\omega^{(d)}(M) = \alpha^{(d)}c/ R_H(M)$ with coefficient
$
\alpha^{(d)} = {(d-3)\over 4\pi} \ln(m_0)
$, where $m_0$ is an integer and 
$\Gamma^{(d-2)}$ is the volume of the unit $d-2$ sphere. 
\end{abstract}
\maketitle

%\clearpage
\section{Introduction}
There is a fairly wide consensus that the mass/entropy spectrum of black holes is discrete rather than continuous. Two methods currently exist for deriving this spectrum. The more fundamental and difficult one is to count black hole states of a fixed energy within some microscopic theory of quantum gravity in order to compute the statistical mechanical entropy. This has been done with some success in string theory\cite{string}(for extremal and near extremal black holes), and in loop quantum gravity\cite{loop}. The second method consists of quantizing (usually within a mini-superspace approach) certain dynamical modes of the classical theory and evaluating the quantum spectrum of the Bekenstein-Hawking entropy, $S_{BH}$, which by definition is one quarter of the horizon area expressed in Planck units. 

 Bekenstein was the first to claim on heuristic grounds that black hole entropy is an adiabatic invariant with an equally spaced quantum spectrum\cite{beken1}. Bekenstein and Mukhanov\cite{beken2} argued that in order for this entropy to have a statistical mechanical interpretation, it must correspond to the logarithm of an
integer: $S_{BH}=k\ln(\Omega)$, with $\Omega =2^n$ being the most natural choice. There have been many attempts to derive this entropy spectrum directly from the dynamical modes of the classical
 theory\cite{kastrup,louko,barv1,barv2,other,joey}. Discrete spectra arise in quantum mechanics in the presence of a periodicity in the classical system which in turn leads to the existence of an adiabatic invariant
or action variable. Bohr-Sommerfeld quantization implies that this adiabatic invariant has an equally spaced spectrum in the semi-classical limit. From this 
viewpoint, the main problem of black hole quantum mechanics has been to correctly identify the physically relevant period or vibrational frequency. For example, in \cite{barv1}, the period was taken to be proportional to the inverse Hawking temperature motivated by the corresponding periodicity in the Euclidean form of the black hole partition function\cite{kastrup}. As shown generically in \cite{barv2} this leads to the result that the Bekenstein-Hawking entropy is indeed an adiabatic invariant with an equally spaced quantum spectrum.

To date there has been very little known about the direct physical connection between the classical dynamical quantities that give rise to the Bekenstein-Hawking entropy and the corresponding microscopic degrees of freedom of the quantum
black hole. Recently tantalizing evidence for such  a connection has emerged\cite{hod,dreyer1} 
in the context of the quasi-normal modes of the classical theory\cite{quasi}. In particular Hod\cite{hod} assumed an equally spaced area spectrum and used the apparent  existence of a unique quasi-normal mode frequency in the large damping limit to uniquely fix the spacing. Remarkably, the spacing was such as to
allow a statistical mechanical interpretation for the resulting eigenvalues for the Bekenstein-Hawking entropy. Dreyer\cite{dreyer1} also used the large damping quasi-normal mode frequency to fix the   value of the Immirzi parameter, $\gamma$, in loop quantum gravity. This value of $\gamma$ turned out to be precisely the one required to make the loop quantum gravity entropy prediction coincide
with the classical Bekenstein-Hawking entropy.

In the present paper we will first use these observation about the quasi-normal mode frequency to derive the general form in the semi-classical limit of the Bekenstein-Hawking entropy spectrum for $d$-dimensional spherically symmetric
black holes. Note that Hod\cite{hod} assumed an equally spaced area spectrum, whereas we will show how it arises as a consequence of taking the large damping quasi-normal mode frequency seriously in the context of black hole dynamics. Moreover, we argue that this analysis allows us to 
predict the $d$-dimensional large damping quasi-normal mode frequency up to a single arbitary integer. If this prediction is confirmed by numerical calculations, then the entropy spectrum could be used as a test for the viability of any proposed microscopic theory of quantum gravity.

  We will first derive the entropy spectrum for $4-d$ black holes in a way that is not tied to any particular microscopic theory of quantum gravity. In the next section we will examine the consequences of this
interpretation for higher dimensional, spherically symmetric black holes. The final section contains conclusions.

\section{4-D Black Holes}
We start from the observation that for a Schwarzschild black hole of mass $M$ and  radius $R_H(M)$, the real part of the quasi-normal mode frequency approaches
a fixed non-zero value in the large damping limit. This value is: 
\beq
\omega_{QNM}(M) = \alpha^{(4)} \left({c\over R_H(M)}\right)
\label{omega4}
\eeq
The functional form $(c/R_H)$ is expected from dimensional arguments:
it is just the inverse of the horizon light transit time. The  coefficient $\alpha^{(4)}$  has been determined
numerically\cite{quasi},\cite{foot}:
\beq
\alpha^{(4)}= 0.04371235 \approx {\ln(3)\over 4\pi}
\eeq 
Following \cite{hod,dreyer1}, we assume that this classical frequency plays an important role in the dynamics of the black hole and is relevant to its quantum properties. In particular, we
consider $\omega_{QNM}(M)$ to be a fundamental vibrational frequency for a black hole of energy $E=Mc^2$. Given a system with energy $E$ and vibrational frequency $\omega(E)$ it is a straightforward exercise to show that the quantity:
\beq
I=\int {dE\over \omega(E)}
\eeq
is an adiabatic invariant, which via Bohr-Sommerfeld quantization has an
equally spaced spectrum in the semi-classical (large $n$) limit:
\beq
I \approx n\hbar
\eeq
By taking $\omega_{QNM}$ seriously in this context, we are lead to the adiabatic invariant:
\beq
I(E) = {c^3\over 2G \alpha^{(4)} }\int {dE\, E}= {\hbar\over 4\pi\alpha^{(4)}} S_{BH}(E) + constant
\eeq
where we have used the fact that $R_H= 2GM/c^2= 2GE/c^4$ and the definition
of the Bekenstein Hawking entropy $S_{BH} = \pi R_H^2/(\hbar G)$. Bohr-Sommerfeld quantization then implies that the entropy spectrum is equally spaced, with
coefficient determined by $\alpha^{(4)}$:
\beq
S_{BH} = 4\pi \alpha^{(4)} n = \ln(3) n
\eeq
The miracle is that the numerical value of $\alpha^{(4)}$ allows
a statistical mechanical interpretation for this entropy. That is, $S_{BH}$ is the logarithm of an integer, which can be interpreted as the degeneracy of quantum states:
\beq
\Omega(E)= \exp(S_{BH}) = 3^n
\label{omega1}
\eeq
It is important to note that this  prediction for the degeneracy of states comes directly from the postulated physical interpretation of
the quasi-normal mode $\omega_{QNM}$. There are no free parameters once $\alpha$ is ``measured'' numerically. It therefore provides in principle a constraint
that any viable quantum theory of gravity must satisfy. Eq.(\ref{omega1}) is similar in form to the degeneracy $\Omega=2^n$ advocated by Bekenstein and Mukhanov\cite{beken2}. Moreover, with suitable assumptions about the gauge group, loop quantum gravity is able to predict precisely this degeneracy of states\cite{dreyer1}.
\section{$d$-Dimensional Schwarzschild Solution}
We now extend the above considerations to spherically symmetric black holes
in $d$-dimensions (the so-called Schwarzschild-Tangherlini black holes\cite{tangherlini}). The solution for the metric, expressed in terms of the ADM energy $E$ is:
\beq
ds^2=-\left(1-{16\pi G^{(d)} E\over (d-2)\Gamma^{(d-2)} c^4 r^{d-3}}\right) + 
   \left(1-{16 \pi G^{(d)} E\over (d-2)\Gamma^{(d-2)} c^4 r^{d-3}}\right)^{-1}dr^2 + r^2 d\Omega^{(d-2)}
\label{tangher_soln}
\eeq
where $G^{(d)}$ is the $d$-dimensional Newton constant, $d\Omega^{(d-2)}$ is the volume element on a unit $(d-2)$-sphere and $\Gamma^{(d-2)}$ is the volume of the unit $(d-2)$-sphere. The corresponding horizon radius is:
\beq
R_H(E)= 
     \left({16\pi G^{(d)} E\over (d-2)\Gamma^{(d-2)} c^4} \right)^{1\over d-3}
\eeq
and the associated Bekenstein-Hawking entropy is:
\beq
S_{BH}(E) = {1\over4}\left({c^3 \Gamma^{(d-2)} R^{d-2}_H(E)\over \hbar G^{(d)}}
\right)
\label{entropy}
\eeq
By applying the first law of thermodynamics, $dS = dE/T_{BH}$ one can deduce
the general expression for the Bekenstein-Hawking temperature:
\beq
T_{BH}(E)= {\hbar c (d-3)\over 4\pi R_H(E)}
\label{temperature}
\eeq

We now assume that, as in 4-d, there is a unique quasinormal mode frequency in the large damping limit whose value on dimensional grounds takes the general form:
\beq
\omega_{QNM}(E)= \alpha^{(d)}{c\over R_H(E)}
\label{omega2}
\eeq
Note that the existence of such a unique limit in dimensions other than 4 has not been established, so that the coefficient is as yet unkown. 
The same semi-classical argument as in the previous section suggests the existence of an adiabatic invariant associated with $d$-dimensional black holes:
\bea
I^{(d)}(E)&=& \int {dE\over \omega_{QNM}(E)}\nonum
 &=&{\hbar (d-3)\over 4\pi \alpha^{(d)}}\int {dE\over T_{BH}(E)}\nonum
  &=&{\hbar (d-3)\over 4\pi \alpha^{(d)}} S_{BH}(E)
\eea
where we have used (\ref{temperature}) and (\ref{omega2}) to replace $\omega$ in the denominator by the
Bekenstein-Hawking temperature, and the first law to arrive at the final expression in terms of the entropy.
Bohr-Sommerfeld quantization of the adiabatic invariant $I=n\hbar$ thus gives rise to an equally spaced entropy spectrum in the semi-classical limit:
\beq
S_{BH}(E)= {4\pi \alpha\over (d-3)} n
\eeq
It is important to remember that it is not merely a numerical coincidence that the Bekenstein-Hawking entropy emerges
as the adiabatic invariant associated with the quasi-normal mode vibrational frequency. Dimensional arguments suggest that for Scharzschild black holes the frequency is proportional to $c/R_H$, which in turn is proportional to the Hawking temperature of the black hole (in any spacetime dimension). Thus the adiabatic invariant $I$ is generically:
\beq
I \propto \int {dE\over T_{BH}}
\eeq
As first argued in \cite{barv2}, it is therefore a direct consequence of the first law of thermodynamics that the entropy is an adiabatic invariant with an equally spaced quantum spectrum. An interesting question is whether this relationship also holds for charged, or rotating black holes.

If one makes the additional assumption that the Bekenstein-Hawking entropy is actually the statistical mechanical
entropy of the black hole states associated with the microscopic quantum gravity theory, then the quantity $\Omega(E)=\exp(S_{BH})=:(m_0)^n$ must be an integer.
This in turn suggests the following form for the coefficient of the quasi-normal mode frequency:
\beq
\alpha^{(d)}= {(d-3)\over 4\pi}\ln(m_0)
\eeq
where $m_0$ is an unspecified integer that depends on the microscopic theory.
\section{Conclusions}
We have argued as in \cite{hod,dreyer1}, that the unique large damping quasi-normal mode of spherically symmetric black holes is a fundamental frequency associated with the dynamical system that can be used to gain semi-classical information about the entropy spectrum of generic black holes. This conjecture leads directly to the result that Bekenstein-Hawking entropy is an adiabatic invariant with an equally spaced quantum spectrum, with uniquely fixed numerical coefficient. As shown by Dreyer\cite{dreyer1} the resulting spectrum is  consistent
with the predictions of loop quantum gravity in 4-$d$ providing the Immirzi parameter is chosen judiciously. Given our higher dimensional results, the validity of the underlying conjecture can be further tested in principle by calculating the coefficients, $\alpha^{(d)}$, of the large damping quasi-normal modes in higher dimensions. This calculation is currently in progress.

Finally, we end with a speculation. One calculation that directly relates the classical dynamics of black holes to a statistical mechanical interpretation of the Bekenstein-Hawking entropy is due to Carlip\cite{carlip}. In this approach, the huge black hole entropy is associated with a degeneracy of states arising fromexistence of an asymptotic conformal symmetry at the horizon. The present work seems to suggest a possible connection classically between the unique quasi-normal mode frequency in the large damping limit, and this asymptotic conformal symmetry. 
In particular, the uniqueness of the quasi-normal mode frequencies in the large damping limit may somehow be related to the conformal symmetry near the horizon. The fact that these frequencies are directly connected to the  
statistical mechanical entropy seems adds weight to this conjecture.

%%%%%%%%%%%%%%%%%%%%%%%%%%%%%%%%%%%%%%%%%%%%%%%%%%%%%%%%%%%%%%%%%%%%%%%%%%%

\vspace{.4cm}
\noindent
{\bf Acknowledgements}

\noindent
I am grateful to O. Dreyer for sharing with me his insights regarding quasi-normal modes. I also thank the Perimeter Institute for its hospitality and R. Epp and S. Das
for helpful conversations.
This work was supported in part by the Natural Sciences and
Engineering Research Council of Canada.

%%%%%%%%%%%%%%%%%%%%%%%%%%%%%%%%%%%%%%%%%%%%%%%%%%%%%%%%%%%%%%%%%%%%%%%%%%%


\begin{thebibliography}{99}
\bibitem{foot}The coefficient in Dreyer's paper differs
by a factor of two since it is expressed in terms of black hole mass rather than radius.
\end{thebibliography}
\end{document}